\newif\ifUsenix\Usenixfalse 
\newif\ifAnon\Anonfalse 

\ifUsenix
  \documentclass[letterpaper,twocolumn,10pt]{article}
  \usepackage{usenix2019_v3}
  \usepackage{authblk}
\else
  \documentclass[conference]{IEEEtran}
\fi
\usepackage{tugraz_defaults}
\usepackage{subfigure}

\usepackage{amsmath}
\usepackage{amssymb}
\usepackage{mathtools}

\usepackage{multirow}
\newcolumntype{P}[1]{>{\centering\arraybackslash}p{#1}}
\newcolumntype{M}[1]{>{\centering\arraybackslash}m{#1}}
\usepackage{tablefootnote}

\makeatletter
\newcommand\footnoteref[1]{\protected@xdef\@thefnmark{\ref{#1}}\@footnotemark}
\makeatother

\usepackage{hyperref}
    \definecolor{linkcolor}{rgb}{0,0.1,0.35}
    \definecolor{citecolor}{rgb}{0,0.4,0}
    \definecolor{urlcolor}{rgb}{0,0,0.65}
    \hypersetup{pdfpagemode=UseNone,pdfstartview=FitH,colorlinks=true, linkcolor=linkcolor, urlcolor=urlcolor, citecolor=citecolor}

\definecolor{dkgreen}{rgb}{0,0.6,0}
\definecolor{gray}{rgb}{0.5,0.5,0.5}
\definecolor{mauve}{rgb}{0.58,0,0.82}
\usepackage[toc,acronym,nowarn]{glossaries}

\newacronym{OS}{OS}{Operating System}
\newacronym{SE}{SE}{Secure Element}
\newacronym{NFC}{NFC}{Near Field Communication}
\newacronym{BLE}{BLE}{Bluetooth Low Energy}
\newacronym{RIL}{RIL}{Radio Interface Layer}
\newacronym{SELinux}{SELinux}{Security-Enhanced Linux}
\newacronym{LSM}{LSM}{Linux Security Modules}
\newacronym{TCB}{TCB}{Trusted Computing Base}
\newacronym{TEE}{TEE}{Trusted Execution Environment}
\newacronym{MAC}{MAC}{Message Authentication Code}
\newacronym{IPC}{IPC}{Inter-Process Communication}
\newacronym{ICC}{ICC}{Inter-Container Communication}
\newacronym{cgroups}{cgroups}{control groups}
\newacronym{CA}{CA}{Certificate Authority}
\newacronym{PKI}{PKI}{Public Key Infrastructure}
\newacronym{MITM}{MITM}{Man-In-The-Middle}
\newacronym{BYOD}{BYOD}{Bring-Your-Own-Device}
\newacronym{CM}{CM}{Container Management}
\newacronym{SM}{SM}{Security Management}
\newacronym{HAL}{HAL}{Hardware Abstraction Layer}
\newacronym{TLS}{TLS}{Transport Layer Security}
\newacronym{C2C}{C2C}{Container To Container}
\newacronym{Protobuf}{Protobuf}{Protocol Buffers}
\newacronym{SDO}{SDO}{Sensitive Data Object}
\newacronym{VMA}{VMA}{Virtual Memory Area}
\newacronym{PGD}{PGD}{Page Global Directory}
\newacronym{PUD}{PUD}{Page Upper Directory}
\newacronym{PMD}{PMD}{Page Middle Directory}
\newacronym[firstplural=Page Table Entries (PTEs)]{PTE}{PTE}{Page Table Entry}
\newacronym{COW}{COW}{Copy-On-Write}
\newacronym{IV}{IV}{Initialization Vector}
\newacronym{ESSIV}{ESSIV}{Encrypted Salt-Sector Initialization Vector}
\newacronym{KSM}{KSM}{Kernel Samepage Merging}
\newacronym{JIT}{JIT}{Just-In-Time}
\newacronym{DMA}{DMA}{Direct Memory Access}
\newacronym{FDE}{FDE}{Full Disk Encryption}
\newacronym{AS}{AS}{Address Space}
\newacronym{GCM}{GCM}{Google Cloud Messaging}
\newacronym{TPM}{TPM}{Trusted Platform Module}
\newacronym{JTAG}{JTAG}{Joint Test Action Group}
\newacronym{LUKS}{LUKS}{Linux Unified Key Setup}
\newacronym{VPN}{VPN}{Virtual Private Network}
\newacronym{PBKDF2}{PBKDF2}{Password-Based Key Derivation Function 2}
\newacronym{KVM}{KVM}{Kernel-based Virtual Machine}
\newacronym{VM}{VM}{Virtual Machine}
\newacronym{firstplural=Virtual Machines}{VM}{VM}
\newacronym{HV}{HV}{Hypervisor}
\newacronym{SEV}{SEV}{Secure Encrypted Virtualization}
\newacronym{SME}{SME}{Secure Memory Encryption}
\newacronym{RMP}{RMP}{Reverse Map Table}
\newacronym{TSME}{TSME}{Transparent \gls{SME}}
\newacronym{SP}{SP}{Secure Processor}
\newacronym[firstplural=Guest Virtual Addresses (GVAs)]{GVA}{GVA}{Guest Virtual Address}
\newacronym[firstplural=Guest Physical Addresses (GPAs)]{GPA}{GPA}{Guest Physical Address}
\newacronym[firstplural=Host Physical Addresses (HPAs)]{HPA}{HPA}{Host Physical Address}
\newacronym{GPT}{GPT}{Guest Page Table}
\newacronym{HPT}{HPT}{Host Page Table}
\newacronym{TLB}{TLB}{Translation Lookaside Buffer}
\newacronym{PoC}{PoC}{Proof of Concept}
\newacronym{ORAM}{ORAM}{Oblivious RAM}
\newacronym{SEV-ES}{SEV-ES}{SEV Encrypted State}
\newacronym{VMCB}{VMCB}{Virtual Machine Control Block}
\newacronym{VMSA}{VMSA}{Virtual Machine State Save Area}
\newacronym{TIK}{TIK}{Transport Integrity Key}
\newacronym{TEK}{TEK}{Transport Encryption Key}
\newacronym{SLAT}{SLAT}{Second Level Address Translation}
\newacronym[firstplural=Nested Page Tables (NPTs)]{NPT}{NPT}{Nested Page Table}
\newacronym{SSH}{SSH}{Secure Shell}
\newacronym{RSA}{RSA}{Rivest–Shamir–Adleman}
\newacronym{ECDHE}{ECDHE}{Elliptic-Curve Diffie-Hellman Ephemeral}
\newacronym{AES}{AES}{Advanced Encryption Standard}
\newacronym{ECB}{ECB}{Electronic Codebook Mode}
\newacronym{OOM}{OOM}{Out Of Memory}
\newacronym{TME}{TME}{Total Memory Encryption}
\newacronym{MKTME}{MKTME}{Multi-Key Total Memory Encryption}
\newacronym{VMI}{VMI}{Virtual Machine Introspection}
\newacronym{MAD}{MAD}{Median Absolute Deviation}
\newacronym{HSM}{HSM}{Hardware Security Module}
\newacronym{AE}{AE}{Automatic Exit}
\newacronym{NAE}{NAE}{Non-Automatic Exit}
\newacronym{AES-NI}{AES-NI}{AES New Instructions}
\newacronym{NIC}{NIC}{Network Interface Card}
\newacronym{NMI}{NMI}{Non-Maskable Interrupt}
\newacronym{MTU}{MTU}{Maximum Transmission Unit}
\newacronym[firstplural=Virtual Addresses]{VA}{VA}{Virtual Address}
\newacronym[firstplural=Physical Addresses]{PA}{PA}{Physical Address}
\newacronym{GFN}{GFN}{Guest Frame Number}
\newacronym{IOMMU}{IOMMU}{I/O Memory Management Unit}
\newacronym{AISE}{AISE}{Address Independent Seed Encryption}
\newacronym{MT}{MT}{Merkle Tree}
\newacronym{BMT}{BMT}{Bonsai Merkle Tree}
\newacronym{LPID}{LPID}{Located Page IDentifier}
\newacronym{CB}{CB}{Counter Block}
\newacronym{PRD}{PRD}{Page Root Directory}
\newacronym{SWIOTLB}{SWIOTLB}{Software I/O Translation Buffer}
\newacronym{ASID}{ASID}{Address Space Identifier}
\newacronym{vCPU}{vCPU}{virtual CPU}
\newacronym{KEK}{KEK}{Key Encryption Key}
\newacronym{KIK}{KIK}{Key Integrity Key}
\newacronym{VEK}{VEK}{VM Encryption Key}
\newacronym{GCTX}{GCTX}{Guest Context}
\newacronym{GHCB}{GHCB}{Guest Hypervisor Communication Block}
\newacronym{XEX}{XEX}{Xor-Encrypt-Xor}
\newacronym{XE}{XE}{Xor-Encrypt}
\newacronym{SGX}{SGX}{Software Guard Extensions}
\newacronym{KASLR}{KASLR}{Kernel Address Space Layout Randomization}
\newacronym{VC}{\texttt{\#VC}}{VMM Communication Exception}
\newacronym{IBS}{IBS}{Instruction Based Sampling}
\newacronym{SEV-SNP}{SEV-SNP}{SEV Secure Nested Paging}
\newacronym{CCA}{CCA}{Chosen Ciphertext Attack}
\newacronym{CPA}{CPA}{Chosen Plaintext Attack}
\newacronym{ROP}{ROP}{Return-oriented Programming}
\newacronym{PAE}{PAE}{Physical Address Extension}
\newacronym{MMIO}{MMIO}{Memory Mapped I/O}
\newacronym{OVMF}{OVMF}{Open Virtual Machine Firmware}

\makeglossaries
\glsdisablehyper

\usepackage[normalem]{ulem}

\newcommand{\ldata}[0]{\texttt{LAUNCH\_UPDATE\_DATA}}
\newcommand{\lvmsa}[0]{\texttt{LAUNCH\_UPDATE\_VMSA}}
\newcommand{\lupdateStar}[0]{\texttt{LAUNCH\_UPDATE\_*}}
\newcommand{\lsecret}[0]{\texttt{LAUNCH\_SECRET}}
\newcommand{\lfinish}[0]{\texttt{LAUNCH\_FINISH}}
\newcommand{\lmeasure}[0]{\texttt{LAUNCH\_MEASURE}}
\newcommand{\lstart}[0]{\texttt{LAUNCH\_START}}
\newcommand{\pdhExport}[0]{\texttt{PDH\_CERT\_EXPORT}}
\newcommand{\suninit}[0]{\texttt{UNINIT}}
\newcommand{\slupdate}[0]{\texttt{LUPDATE}}
\newcommand{\slsecret}[0]{\texttt{LSECRET}}
\newcommand{\srunning}[0]{\texttt{RUNNING}}

\DeclareMathOperator{\hash}{hash}
\DeclareMathOperator{\HMAC}{HMAC}
\newcommand{\concat}[0]{\,\|\,}

\begin{document}

\title{
	\Large \bf undeSErVed trust: Exploiting Permutation-Agnostic Remote Attestation
}

\ifAnon  
  \author{Anonymous Submission}
\else
  \ifUsenix
    \author[1]{Luca Wilke}
    \author[1]{Jan Wichelmann}
    \author[1]{Florian Sieck}
    \author[1]{Thomas Eisenbarth}
    \affil[1]{University of L\"ubeck, L\"ubeck, Germany}
  \else
    \author{
      \IEEEauthorblockN{
        Luca Wilke, 
        Jan Wichelmann, 
        Florian Sieck, 
        Thomas Eisenbarth 
      }
      \IEEEauthorblockA{University of L\"ubeck, Germany\\
      \{l.wilke,j.wichelmann,florian.sieck,thomas.eisenbarth\}@uni-luebeck.de}
    }
  \fi
\fi

\maketitle
\thispagestyle{plain}
\pagestyle{plain}

\begin{abstract}
The ongoing trend of moving data and computation to the cloud is met with concerns regarding privacy and protection of intellectual property. Cloud Service Providers (CSP) must be fully trusted to not tamper with or disclose processed data, hampering adoption of cloud services for many sensitive or critical applications. As a result, CSPs and CPU manufacturers are rushing to find solutions for secure and trustworthy outsourced computation in the Cloud.
While enclaves, like Intel SGX, are strongly limited in terms of throughput and size,
AMD's Secure Encrypted Virtualization (SEV) offers hardware support for transparently protecting code and data of entire VMs, thus removing the performance, memory and software adaption barriers of enclaves. Through attestation of boot code integrity and means for securely transferring secrets into an encrypted VM, CSPs are effectively removed from the list of trusted entities.
There have been several attacks on the security of SEV, by abusing I/O channels to encrypt and decrypt data, or by moving encrypted code blocks at runtime. Yet, none of these attacks have targeted the attestation protocol, the core of the secure computing environment created by SEV. 
We show that the current attestation mechanism of Zen 1 and Zen 2 architectures has a significant flaw, allowing us to manipulate the loaded code without affecting the attestation outcome.  An attacker may abuse this weakness to inject arbitrary code at startup---and thus take control over the entire VM execution, without any indication to the VM's owner. Our attack primitives allow the attacker to do extensive modifications to the bootloader and the operating system, like injecting spy code or extracting secret data. We present a full end-to-end attack, from the initial exploit to leaking the key of the encrypted disk image during boot, giving the attacker unthrottled access to all of the VM's persistent data.
\end{abstract}

\section{Introduction}

An increasing number of software applications, from enterprise management software to messengers used in nearly everyone's daily life, rely on storing information and performing computations in the cloud. Solutions are moved from local, trusted environments to the data centers of big cloud service providers, and are now running in untrusted environments under control of a third party---in order to save costs, reduce management effort and to improve scalability.

The loss of trust comes with significant challenges for services such as banking, private secure messaging or health services, which require strict isolation and confidentiality to ensure the safety of their assets and to comply with data privacy laws: Computing resources in the cloud are often shared, which in case of broken isolation does allow co-located users to spy on each other \cite{inci2016cache, liu2015last, zhang2012cross}. Another concern is the security of the cloud service provider's systems themselves, where internal or external attackers may leverage elevated privileges for extracting private data. 

In order to deliver isolated, confidential and authenticated execution and processing of data in an otherwise untrusted setting, processor vendors added hardware features to build a root-of-trust and ensure confidential computing in a local \gls{TEE}. One example is AMD SEV \cite{AMD2020API, AMD2020architecture, kaplan2016amd}, which allows to run VMs confidentially and isolated from their hypervisor. AMD added new features to SEV with every generation of its processor architecture. In 2017, The first generation of EPYC processors (Zen) came with the initial version of SEV. The second generation (Zen 2) added an encrypted state for context switches with SEV-ES \cite{sev-es} and was released in 2019. The newest addition, SEV-SNP \cite{sev-snp}, will be available on the third generation of EPYC processors (Zen 3), which are set to be released in early 2021. Intel TDX~\cite{inteltdx2020} aims to provide a similar solution, but is only available as a concept as of writing this work. With Intel \gls{SGX} \cite{sgxdeveloperguide, anati2013innovative, costan2016intel}, Intel offers an established \gls{TEE} which enables software vendors to run smaller programs in isolated enclaves. 
All of these solutions provide memory encryption during execution, and attestation of the software loaded into the \gls{TEE}.

Recently, cloud service providers like Microsoft and Google started to offer confidential computing environments which isolate the customer's software using Intel SGX \cite{azureconfidcomputing} or AMD SEV \cite{googleconfidcomputing}. Popular examples, like the secure private messenger Signal, are already using these technologies to protect the sensitive data of their customers \cite{gaazureconfcomp}.
Moreover, open source solutions enable simple development and deployment of software for \gls{TEE}s \cite{enraxgithub, asylogithub, arnautov2016scone}.

A fundamental challenge for \glspl{TEE} is having to guarantee their promises against attackers with system level privileges, resulting in a large variety of attacks \cite{DBLP:journals/corr/abs-2010-07094, DBLP:conf/sp/WilkeWM020, buhren2019insecure, moghimi2020copycat, van2020lvi, van2018foreshadow}. In this work, we extend the arsenal of attacks against \gls{TEE}s and in particular against AMD SEV, with an attack targeting and circumventing its very core of trust, the remote attestation.

Remote attestation allows the owner of a software, which runs in a confidential or trusted execution environment, to verify the initial integrity and authenticity of the software loaded into the \gls{TEE}, which afterwards is preserved at runtime by the properties of the \gls{TEE}. Generally, remote attestation works by creating a signed measurement, usually a hash, of the initially loaded application through the trusted hardware and sending this measurement to the software owner for verification. In case of AMD SEV, the trusted hardware is an additional on-chip co-processor called \gls{SP}, which cannot be externally controlled. 

\subsection{Our Contribution}
If the attestation process, however, is broken, the isolation and confidentiality guarantees of AMD SEV are inconsequential as the software owner cannot be sure whether their intended software was loaded or whether an attacker manipulated it during startup.

\noindent In this work, we
\begin{itemize}
    \item show that the measurement used in AMD SEV's attestation is block permutation-agnostic, meaning that changing the order of measured memory blocks does not affect the attestation outcome, and thus allows the attacker to modify the execution flow without detection by the VM's owner;
    \item construct an universal attack primitive, which reorders the measured blocks of an initially loaded UEFI and sets up a \gls{ROP} chain to load and execute arbitrary code;
    \item demonstrate a full end-to-end attack which leaks the key of an encrypted disk image, and gives the attacker full control over the VM's operating system;
    \item propose several countermeasures and discuss why the underlying problem ultimately cannot be solved under \gls{SEV-ES}.
\end{itemize}

\subsection{Attack Overview}
The attack described in this work targets the measurement of the initially loaded binary during startup of an SEV-ES-protected \gls{VM} (guest). When the \gls{VM} is started, the hypervisor instructs the AMD SEV secure processor to load the initial binary, e.g. an \gls{OVMF} UEFI binary, into encrypted memory and calculate a measurement of the initial VM content using the \ldata~and \lmeasure~commands. We find that the initial binary can be split into blocks as small as 16 bytes, which we are able to load in an arbitrary order using \ldata~, while still getting the same measurement when calling \lmeasure. This allows us to construct our own execution flow, which we use for redirecting the stack pointer to an unencrypted shared page. Consequently, we leverage this control over the stack to mount a \gls{ROP} attack, allowing us to write arbitrary code and data into the encrypted VM's memory. We use the injected code to leak the protected secret values which have been provided by the guest owner. As our meddling with the block ordering does not change the launch measurement, AMD SEV's remote attestation will succeed and the guest owner will be unaware of our changes to their VM's execution flow.

\subsection{Responsible Disclosure}
We responsibly disclosed our findings to AMD via Email on January 19th, 2021. AMD requested an embargo until
May 11, 2021 and provided us with the following statement:
``AMD has assigned CVE-2021-26311 for this issue and provided mitigations in the SEV-SNP feature available for enablement 3rd Gen AMD EPYC™ processors. AMD appreciates the coordination efforts made by the research team.''.

\section{Background}

\subsection{AMD SEV} \label{text:sev-background}

In 2016, AMD introduced their \gls{SME} and \gls{SEV} technologies~\cite{kaplan2016amd}, which were implemented only in 2017 with the first generation of EPYC processors (Zen 1).
\gls{SME} offers hardware-based encryption of RAM content. 
The memory encryption key is managed by the \gls{SP}, an ARM-based co-processor, and is thus never accessible by system software. The encryption/decryption takes place directly in the on-die memory controllers. Each page table entry has a special status bit, which controls whether the associated page is encrypted or not. The whitepaper~\cite{kaplan2016amd} does not explain the mode of operation in detail, but only states that AES with an 128-bit key and a physical address-based tweak is used.
In \cite{du2017secure,DBLP:conf/sp/WilkeWM020} it is shown that early versions use the \gls{XE} or \gls{XEX} encryption mode with static, low entropy tweak values, while later versions use stronger, randomized tweak values. In addition, none of the encryption modes offer integrity protection.

While \gls{SME} uses the same key for all memory pages, \gls{SEV} adds the ability to encrypt
the memory content of \glspl{VM} with different keys, that are only known to the \gls{SP} but not to the \gls{HV}, preventing a malicious \gls{HV} from directly reading the memory content of its guests. However, \glspl{VM} can also share pages with the \gls{HV}. In addition, the \gls{SP} offers an API to the \gls{HV} to manage the \gls{SEV}-protected \glspl{VM}. This includes a mechanism to attest the initially loaded code of the \gls{VM} and a mechanism to securely move secrets into the \gls{VM}.

\noindent\textbf{\gls{SEV-ES}} was introduced by AMD in 2017 and implemented in 2019 with the second generation of EPYC processors. It addresses
one major remaining attack surface of \gls{SEV}: The unencrypted \gls{VMCB}, a data structure storing certain configuration bits as well the \gls{VM}'s register values on context switches. 
Certain sensitive parts of the \gls{VMCB} were moved to a substructure called \gls{VMSA} that is encrypted and integrity protected on context switches to the \gls{HV}, and thus prevents an attacker from inferring or modifying a VM's state during context switches.

However, there are also several instructions that need interaction with the \gls{HV}, like \texttt{cpuid}, which previously shared and received data with/from the \gls{HV} via the \gls{VM}'s registers. To enable this with \gls{SEV-ES}, AMD introduced a new communication mechanism between \gls{HV} and \gls{VM}, consisting of the \gls{GHCB} and a new exception called \gls{VC}. The \gls{GHCB} is simply a shared page, that gets setup by the \gls{VM}. Instructions that require data sharing with the \gls{HV} cause a \gls{VC}, allowing a \gls{VM} exception handler to share the data via the \gls{GHCB}.

\noindent\textbf{\gls{SEV-SNP}} aims to address several remaining issues, like remapping attacks due to the \gls{HV}'s control over the nested page tables, or attacks on the missing integrity protection.
It was announced by AMD in January 2020~\cite{sev-snp} and will only be available on the 3rd gen EPYC processors, that are set to be released after the submission deadline. The most important change is the introduction of an additional page table called \gls{RMP}, to which the \gls{HV} only has mediated access.
The \gls{RMP} aims to ensure a one-to-one mapping between \glspl{GPA} and \glspl{HPA}, and will prevent the \gls{HV} from writing to \gls{VM} memory, mitigating problems arising due to the lacking integrity protection.

\subsection{Booting physical and virtual environments}
When booting a physical system, the CPU is in a well-defined state, but completely unaware of its environment. Its program counter is set to start execution at a fixed address, which points to a FLASH or EPROM. Firmware is loaded from this start position and is responsible for initializing the memory controller, creating a memory mapping for RAM, and configuring I/O peripherals. On modern computers, this firmware usually is UEFI-based, which is platform specific and performing the aforementioned tasks. The advantage of UEFI compared to legacy BIOS is its standardization of the platform initialization procedure~\cite{uefispec}. The UEFI is configurable through variables in a non-volatile memory (e.g. NVRAM), so configuration is persisted over restarts. Additionally, it can provide \emph{secure boot}, which allows to build up a chain of trust from the UEFI to the finally loaded kernel of the OS. In this chain, the UEFI, which contains a set of configurable certificates and keys, forms the root of trust. Every component of the chain verifies its successor before handing over the execution~\cite{uefispec}.

When the UEFI finishes the system configuration, it hands over the control to an EFI binary~\cite{uefispec}, which usually is an OS bootloader, e.g. Grub~\cite{gnugrub}. The bootloader sets up the stage for the OS kernel, loads the kernel into memory and calls its main method. However, an EFI binary does not necessarily have to be a bootloader.

In case of booting a virtual environment, e.g. with QEMU~\cite{qemuhome}, the process is similar: When the hypervisor starts up the virtualization, it launches an UEFI. For virtual environments, \gls{OVMF}~\cite{ovmfgithub} is a common choice. \gls{OVMF} performs the necessary virtual system configuration and hands over control to a bootloader~\cite{bootqemuperspective}. The bootloader, which is just a regular EFI application, can be provided in different ways. Usually it is expected to be located on a FAT-formatted disk with GUID partition table~\cite{uefispec}, thus requiring the guest owner to provide the hypervisor with a disk image. Another way is to include the bootloader, i.e. the EFI application, into the UEFI volume which also contains the UEFI firmware \cite{sevencbootforovmf}.

\section{SEV(-ES) Guest Launch Process}\label{sec:launch-process}
In this section, we describe the typical workflow required to launch an \gls{SEV}-secured \gls{VM}, as described in~\cite{AMD2020API} and implemented in AMD's patches to the the Linux kernel~\cite{amdSEVESKernelRepo} and QEMU~\cite{amdQemuRepo}. This includes encrypting an initial code image, proving its integrity to the guest owner, and loading secret data without leaking it to the \gls{HV}.

\subsection{Prerequisites}
There are three parties involved in launching a \gls{SEV} \gls{VM}: The guest owner, the
\gls{HV} and the \gls{SP}. The guest owner wants to start a \gls{SEV}-secured \gls{VM}.
The \gls{HV} is typically controlled by the cloud service provider. In order to provide the \gls{SEV} functionality, the HV must interact with the API provided by the \gls{SP}.

\begin{figure}
    \centering
    \includegraphics[width=0.48\textwidth]{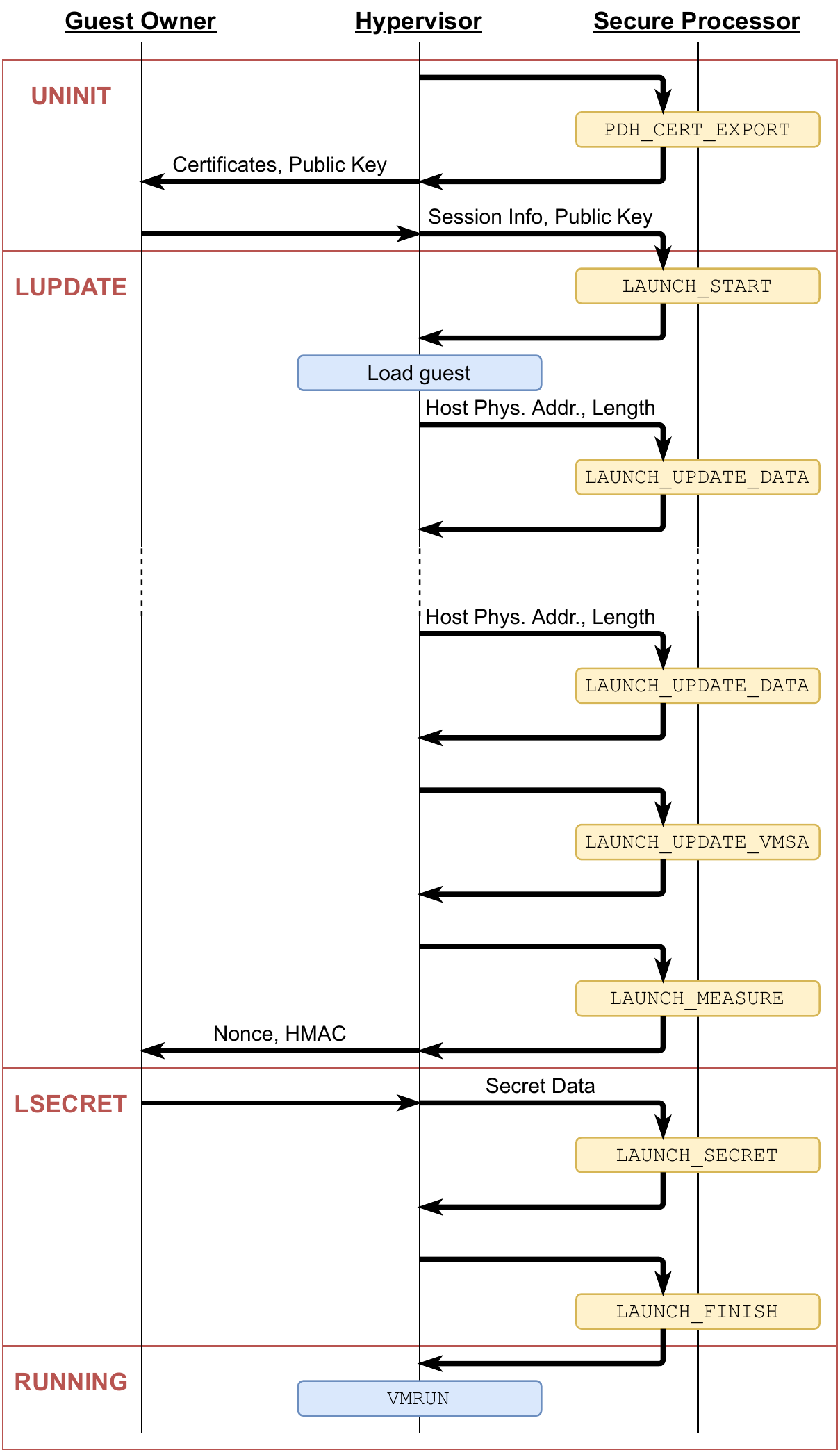}
    \caption{SEV Guest Launch Process. This simplified illustration shows the various states during VM startup and attestation. First, keys are exchanged and a cryptographic session is established. In the \slupdate{} state, the hypervisor loads the guest and then asks the \gls{SP} to encrypt it via repeated \lupdateStar{} commands. The final \lmeasure{} call retrieves a signed hash of the loaded guest data. If the guest owner approves, various secret data can be safely loaded into the VM during the \slsecret{} state. On completion, the hypervisor and the VM transition into the RUNNING state, and the VM is executed.}
    \label{fig:launch}
\end{figure}

The goal of the launch process is to enable the \gls{HV} to prove to the guest owner that the initial content is trustworthy. Furthermore, it enables the guest owner to send secrets, like disk encryption or SSH keys, to the VM in a secure manner. The entire launch process is illustrated in \autoref{fig:launch}.

\begin{table}
    \caption{Overview of the \gls{VM}-specific commands provided by the SP in different states. For brevity, commands that are not relevant for our work were omitted.}
    \begin{tabular}{l|lll}
    State  & Command &  $\rightarrow$ &New State\\
    \hline
    \suninit{} & \lstart{} & $\rightarrow$ &\slupdate{} \\
   \hline
    \slupdate{} & \ldata{} & $\rightarrow$ & \slupdate{} \\
                     & \lvmsa{} & $\rightarrow$ &\slupdate{} \\
                     & \lmeasure{} &$\rightarrow$ &\slsecret{} \\
    \hline
    \slsecret{} & \lsecret{} & $\rightarrow$ & \slsecret{} \\
                     & \lfinish{} & $\rightarrow$ &\srunning{} \\

    \hline
    \srunning{} & (other commands)
    \end{tabular}
    \label{tab:states}
\end{table}

For each \gls{VM}, the \gls{SP} maintains a \gls{GCTX} that, among other values, contains a handle, the \gls{VEK}, the launch digest (LD), and the current state. The \gls{VEK} is a VM-specific key used for memory encryption. The launch digest contains a hash value of the VM contents loaded during the launch measurement phase. The state determines which API commands are usable.

\autoref{tab:states} shows an overview of the states along with the usable commands and the resulting state transitions. We omitted all states and commands related to migrating \glspl{VM} between
different hosts, as we do not use them in this paper. In addition to the \gls{VM}-specific commands, there are several commands which affect the \gls{SP} itself. They are used to update its firmware and to generate or export cryptographic key material.

Before any \gls{VM}-specific commands are issued, the \gls{HV} starts an ECDH key exchange by
issuing the \pdhExport{} command, upon which the \gls{SP} exports a public ECDH key and some certificates. The latter are part of a public key infrastructure, that is ultimately rooted at an AMD controlled key hardcoded into the \gls{SP}. The \gls{HV} then sends this data to the guest owner.

\subsection{\suninit{} state}
A new \gls{VM} assumes \suninit{} as initial state. In order to start the launch process, the guest owner verifies the authenticity of the ECDH key sent by
the \gls{HV}. Then they use their own ECDH key pair to derive the \gls{TEK}, the \gls{TIK} and some other keys used for transport security. Afterwards, they send this data together with a configuration object called \texttt{POLICY} to the \gls{HV}.

Upon receiving the data from the guest owner, the \gls{HV} calls the \lstart{} command, which finalizes the ECDH handshake between guest owner and \gls{SP}. The \gls{SP} can now derive the shared secret and use it to unwrap/verify the received data. Next, it initializes the \gls{GCTX} using the received guest policy and generates a new \gls{VEK}.

\subsection{\slupdate{} state}
In the \slupdate{} state, there are three primary commands: \ldata, \lvmsa~and \lmeasure. The \ldata~command allows the \gls{HV} to specify a guest handle, a 16 byte aligned \gls{HPA} \texttt{PADDR} and a multiple of 16 bytes \texttt{L} as a length. The \gls{SP} will then in-place encrypt the next \texttt{L} bytes starting at \texttt{PADDR} with the \gls{VEK} of the \gls{VM} denoted by the handle. In addition, the launch digest field of the \gls{GCTX} is updated with the plaintext of the encrypted data (see Section \ref{sec:breaking_the_launch_measurement}).
The intention of \ldata~is to encrypt and measure the initial content of the \gls{VM}, such that the \gls{HV} can no longer modify it. Encrypting the initial content is mandatory, since the \gls{VM} initially assumes that all memory accesses are encrypted, so it can only execute the initial code if it has been encrypted beforehand.

The \lvmsa{} command is only applicable to \gls{SEV-ES} \glspl{VM} and works very similar to \ldata{}, except that it can only load 4096 bytes, as it is intended to encrypt the \gls{VMSA}. In addition, it also initializes the \gls{VMCB}. While not enforced, this is intended to be called only once. Again, the launch digest is updated with the loaded data.

The third and final command, \lmeasure{}, generates a launch measurement and transfers the \gls{VM} to the \slsecret{} state. The measurement consists of a 128-bit nonce \texttt{MNONCE} and a 256-bit HMAC \texttt{MEASURE}, that is calculated as follows:
\begin{enumerate}
    \item Replace launch digest (\texttt{LD}) with $\hash(\mathtt{LD})$
    \item Calculate 
    \begin{align*}
        \HMAC(&\mathtt{0x04} \concat \mathtt{API\_MAJOR} \concat \mathtt{API\_MINOR}  \\ 
              &\concat \mathtt{BUILD} \concat \mathtt{POLICY} \concat \mathtt{LD} \\
              &\concat \mathtt{MNONCE}, \mathtt{TIK})
    \end{align*}
    
\end{enumerate}
\texttt{MNONCE} is generated by the \gls{SP}, \texttt{API\_MAJOR} and \texttt{API\_MINOR} and \texttt{BUILD} specify the version of the firmware on the \gls{SP}. \texttt{POLICY} is the configuration structure that was sent by the guest owner in the \suninit{} state.

Next, the \gls{HV} sends the launch measurement to the guest owner, in order to prove that it did not manipulate
the initial content. It is assumed that the guest owner and the \gls{HV}/cloud service provider negotiated the initial content of the \gls{VM}, e.g., that the guest owner stated that they want a specific UEFI version to be loaded. Thus, the guest owner has all the information required to compute the HMAC themselves and compare it to the value they received.

After successfully checking the launch measurement, the guest owner can be sure that the initial memory content matches their specification. Since, on startup, the \gls{VM} treats any memory as encrypted, it is unlikely that the \gls{HV} can achieve any meaningful manipulation of the \gls{VM}'s code and data by tampering with its memory. The only possibility for the \gls{HV} to encrypt data with the \gls{VM}'s key is by using the designated \lupdateStar{} commands, but, as already explained, this has the side effect of updating the launch digest and thus changing the HMAC sent in the attestation report, allowing detection by the guest owner.
As only the \gls{SP} and the guest owner know the \gls{TIK} used to key the HMAC, the \gls{HV} cannot produce valid HMACs itself.

\subsection{\slsecret{} state}
After the \gls{VM} has transitioned into the \slsecret{} state, two commands become available: \lsecret{} and \lfinish{}. The \lsecret{} command again allows to encrypt data with the \gls{VM}'s \gls{VEK}. However, contrary to the previous commands, the data passed to the command now is encrypted with the \gls{TEK} and integrity protected by an HMAC keyed with the \gls{TIK}. Both keys are only known to the \gls{SP} and the guest owner. If the integrity check fails, the command aborts. The guest owner can use this mechanism to safely send confidential data (e.g., disk encryption keys) to the VM, while using the \gls{HV} as a proxy. The \gls{HV} could refuse to relay the data to the \gls{SP}, but it can neither manipulate the data nor call the command with self-generated data, as it does not know the \gls{TIK} needed to pass the HMAC check.

Finally, the \lfinish{} command transitions the \gls{VM} into the \srunning{} state, indicating that the \gls{VM} is ready to be started.
The \lsecret{} and \lfinish{} commands are disabled afterwards.

\section{Attacker Model}
The attacker model is in line with SEV's security model: The attacker controls the hypervisor, and is able to modify arbitrary physical memory and run or pause the \gls{VM} according to their wishes. However, they are not able to read or modify the current register state and program counter of the \gls{VM}, as its state is encrypted using \gls{SEV-ES}.

They know the initially launched code, since that needs to be available in plaintext in order to be loaded into the \gls{VM}. We assume that the attestation is working in so far that the attacker has to actually load and attest the supplied initial code image, and cannot simply replace it with their own. We also assume that the attestation protocol is carried out correctly, such that the guest owner is assured that supposedly the correct image was loaded, and subsequently launches the virtual machine.

The attacker is not able to read or modify encrypted disk images without knowing the corresponding key. Finally, we consider the \gls{SP} itself to be secure.

\section{Exploiting SEV's Permutation Agnostic Launch Measurement}
\label{sec:exploit}
Given the \gls{VM} attestation process laid out in \autoref{sec:launch-process}, we show how an attacker can deviate from the intended startup process in order to make the \gls{VM} execute arbitrary code, which corresponds to a full break of confidentiality and integrity.

In a first step, we show that \gls{SEV}'s launch measurement can be tricked into producing the same measurement for any blockwise permutation of the initial \gls{VM} content.
We illustrate how an attacker can use this flaw to construct an encryption/decryption/code execution gadget, that runs within the \gls{VM}, but does not change its launch measurement and thus cannot be detected by the guest owner.
Finally, in \autoref{sec:attack-scenarios}, we discuss the implications of our attack for
the transition from initially attested code to code residing on a virtual hard disk,
and demonstrate how we can use our attack to leak secrets.

\subsection{Breaking the Launch Measurement}
\label{sec:breaking_the_launch_measurement}
First, we show how a malicious \gls{HV} can abuse a flaw in the launch process to change the semantics of the loaded data without changing the launch measurement.

As described in \autoref{sec:launch-process}, the \gls{HV} uses the \ldata{} command
to load and encrypt the initial memory content of the \gls{VM}.
The command takes a 16-byte aligned
\gls{HPA} \texttt{PADDR} and a multiple of 16 bytes as length \texttt{L}, and then in-place encrypts \texttt{L} bytes starting at \texttt{PADDR} with the \gls{VM}'s \gls{VEK}. In addition, the command updates the launch digest which is later used in the launch measurement.

In our experiments, we observed that the content of the launch digest is neither influenced by the \glspl{HPA} passed to \ldata{}, nor by the used block size and the resulting varying number of calls to the command. Instead, the encrypted data is simply ``appended'' to the launch digest. While the official documentation is unclear at this point, we suspect that the launch digest internally manages a SHA-256 hash state, which is updated each time after a certain amount of data was inserted.

\begin{figure}
    \centering
    \includegraphics[width=0.48\textwidth]{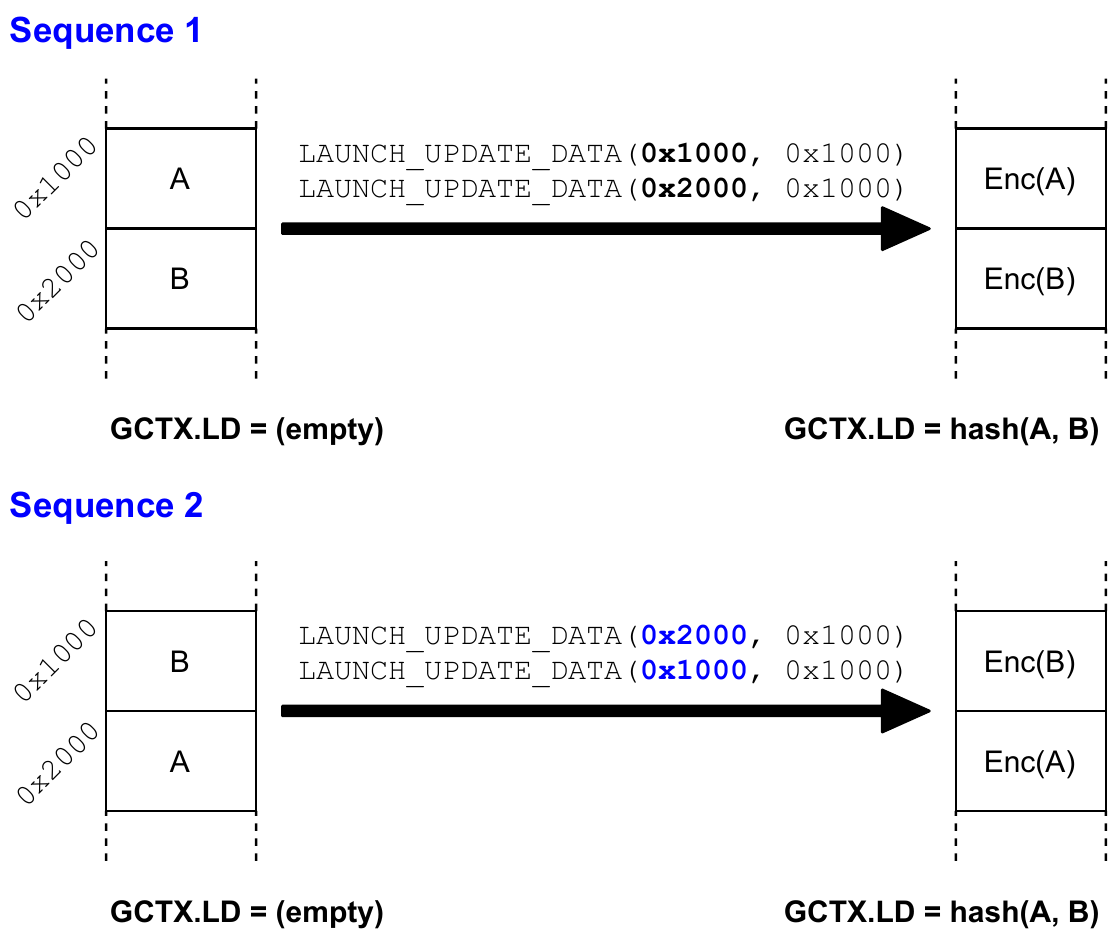}
    \caption{Two encryption sequences, yielding the same launch measurement $\mathrm{GCTX.LD}$ for different orders of memory blocks. In the first sequence, the memory pages A (address \texttt{0x1000}) and B (address \texttt{0x2000}) are encrypted in memory order, i.e., \ldata~is first called for block A, then for block B. In the second sequence, the blocks are swapped in memory: A now resides at address \texttt{0x2000}, while B is at address \texttt{0x1000}. By changing the order of calls to \ldata, we are able to acquire the same value for $\mathrm{GCTX.LD}$ in sequence 2 as for the ``correct'' ordering in sequence 1. The guest owner thus has no means for distinguishing which sequence has been used by the \gls{HV}.}
    \label{fig:reordering}
\end{figure}

This implies that the \gls{HV} can change the memory layout of the loaded data without any impact on the resulting hash value, as long as it makes sure that the order in which the data is passed to \ldata{} matches the original order. The modified ordering is illustrated in \autoref{fig:reordering}.

\subsection{Constructing Malicious Code Gadgets} \label{sec:oracle}
We can now use our observations to construct malicious code gadgets, solely by moving around 16-byte blocks and triggering interaction with the \gls{HV}.

The general idea is very similar to the approach presented in~\cite{DBLP:conf/sp/WilkeWM020}, where the authors leverage control over the first and last bytes of 16-byte blocks to stitch together a sequence of ``payload'' instructions and direct jumps, which they subsequently use to build an encryption oracle within the \gls{VM}. However, we cannot change a block's content here, as this would be detected during attestation.

We first scan the binary of the initial \gls{VM} content, which, in our case, can be split up into 230'000 16-byte blocks, for the instructions that we want to execute
in our gadgets. For this, we are not bound to the originally intended decoding order: As x86 instructions have variable length and are not prefix-free, starting to decode the binary with different offsets can lead to different valid instructions. On the downside, this also means that decoding may fail because it encounters an invalid instruction encoding. To address this, we only look for ``payload'' instructions which reside at the end of a block or are followed by a direct jump, such that we can proceed to the next block.

Finally, we analyze the control flow of the original program to find a location where we can place our gadget, so it is executed at some point during the startup of the \gls{VM}. We also make sure that our changes to the block ordering do not destroy the code needed to boot up the \gls{VM} to the point where our gadget is entered.

\subsection{Encryption/Execution Gadget}
We can now use this block chaining technique to build a code gadget that enables the \gls{HV} to encrypt (and decrypt) arbitrary data with the \gls{VEK} and inject it into the \gls{VM}.

For this, we assume that the \gls{VM} is started with the default \gls{OVMF} UEFI provided by AMD as initial memory content.
Note that the ideas presented here are also applicable to other UEFI implementations that support \gls{SEV}.

The encryption/execution gadget is constructed in two stages:
\begin{enumerate}
    \item Permute the initial \gls{VM} content, creating a gadget that maps the stack
    to an unencrypted shared memory page;
    \item Use the control over the stack to construct a \gls{ROP} gadget that copies data from the unencrypted shared page to encrypted memory and execute it as code. This code may later copy decrypted memory back to the shared page, or can be used to conduct other, more complex attacks.
\end{enumerate}

\begin{figure}
    \centering
    \includegraphics[width=0.45\textwidth]{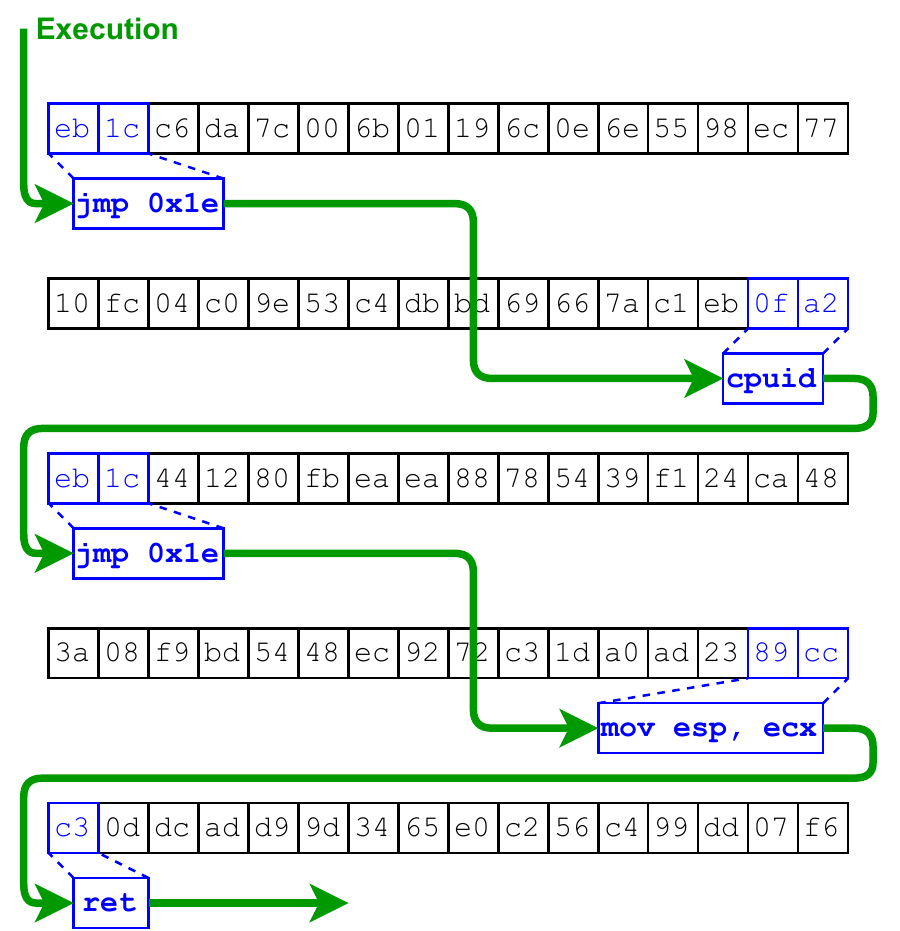}
    \caption{Sequence of 16-byte blocks for setting the stack pointer to a HV-controlled address. The shown 16-byte blocks were taken from various places in the initial code image and moved to an address which is reached by the execution flow. The \texttt{jmp} instructions allow us to chain several blocks and skip potential junk bytes in between. After the stack pointer has been changed, the \texttt{ret} statement reads its return address from HV-controlled memory and thus triggers a \gls{ROP} chain.}
    \label{fig:block-chain}
\end{figure}

\smallskip\noindent\textbf{Stage 1}
In the first stage, we want to set the \gls{VM}’s stack (i.e., its \texttt{rsp} register) 
to an unencrypted memory page, to allow manipulation by the HV. 
Until reaching either long mode or legacy \gls{PAE} mode, the \gls{VM}'s memory
accesses are unconditionally treated as encrypted. Afterwards, the \texttt{C} bit in the page table controls whether a page is accessed in encrypted or unencrypted mode. The only exception are page table walks and instruction fetches, which are always treated as encrypted~\cite[Sec 15.43.4,15.34.5]{AMD2020architecture}.

After startup, \gls{OVMF} quickly progresses to long mode. While constructing its long mode page tables, \gls{OVMF} also sets up a shared page for the \gls{GHCB} protocol~\cite{AMD2020GHCB}, which, under \gls{SEV-ES}, is required to handle the emulation of instructions that need to share data with and/or receive data from the \gls{HV} (c.f. \autoref{text:sev-background}).

To load the address of the shared \gls{GHCB} page into the \texttt{rsp} register, we opted for the following payload instruction sequence:
\begin{itemize}
    \item \texttt{cpuid}\\
    Fills the \texttt{eax}, \texttt{ebx}, \texttt{ecx} and \texttt{edx} registers with 
processor feature information. As shown in previous work~\cite{DBLP:conf/sp/WilkeWM020}, one can abuse
that this instruction is emulated by the \gls{HV} and fill
the \texttt{ecx} register with the virtual address of the shared \gls{GHCB} page.

    \item \texttt{mov esp, ecx}\\
    Updates the stack pointer with the address of the shared page. Note the usage of 32-bit registers: This has the advantage of having a shorter instruction encoding than the 64-bit equivalent, while still being sufficient, as in \gls{OVMF} the virtual address of the shared page is hardcoded to a small constant.
    
    \item \texttt{ret}\\
    Starts the \gls{ROP} chain. As the stack pointer now points to the unencrypted shared page, the \gls{HV} can place arbitrary return addresses (and other values) on the stack, which allows to conduct a classic \gls{ROP} attack.
\end{itemize}

The resulting block chain is illustrated in \autoref{fig:block-chain}.
To ensure the right timing for sending the manipulated \texttt{cpuid} register values, the \gls{HV} simply counts the number of emulated \texttt{cpuid} instructions, which is deterministic in the executed \gls{OVMF} code.

\smallskip\noindent\textbf{Stage 2}
Coming from Stage 1, we now have full \gls{ROP} capabilities.

In order to load additional data into the \gls{VM} and execute it as code, we need
to construct a \gls{ROP} chain that copies data from the unencrypted stack to a memory location 
that is marked as encrypted. Then, we jump to that address using a \texttt{ret} instruction.
This indirect approach of encrypting the code before execution is necessary, since, as mentioned in Stage 1, instruction fetches always assume that the underlying memory is encrypted.
We abuse that, at the time of gadget injection, \gls{OVMF}'s page tables have all pages marked as readable/writable, without having a "no execute" bit. This simplifies our attack, since we do not have to build a \gls{ROP} chain for modifying the page tables first.

\begin{figure}
    \centering
    \includegraphics[width=0.48\textwidth]{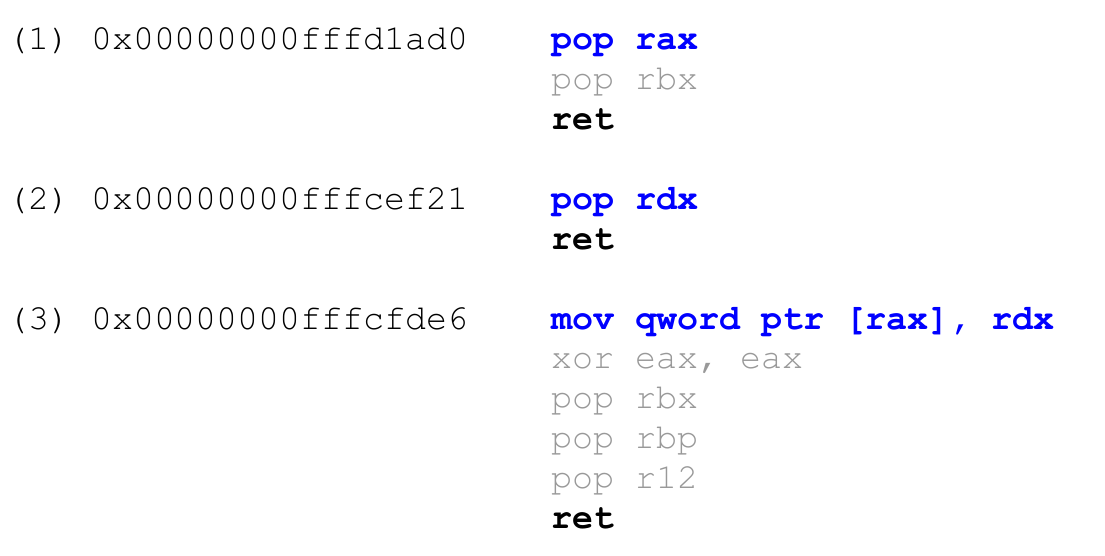}
    \caption{\gls{ROP} chain for writing data to VM memory. The \gls{ROP} chain consists of three gadgets. Each gadget begins with a payload instruction (highlighted blue) and ends with a return statement, potentially with a few other instructions in between. Gadget (1) loads an unencrypted 8-byte address from the stack and writes it into \texttt{rax}. Gadget (2) loads an unencrypted 8-byte value from the stack and writes it into \texttt{rdx}. Finally, gadget (3) stores the value from \texttt{rdx} at the address pointed to by \texttt{rax}. The memory write in (3) triggers encryption of the data stored in \texttt{rdx}, as the address in \texttt{rax} points to encrypted memory.}
    \label{fig:rop-chain}
\end{figure}

Our resulting \gls{ROP} chain is illustrated in \autoref{fig:rop-chain}. It only needs three gadgets, and allows us to write 8 bytes to an arbitrary 64-bit address. We can then reuse the chain to write complex new code into the \gls{VM}.
In \autoref{sec:attack-scenarios}, we show how this code can be used to leak disk encryption keys.

In summary, we have seen that \gls{SEV}'s launch measurement mechanism is flawed, as it only attests that an arbitrary 16-byte granular permutation of the initial data has been loaded. Next, we have demonstrated the creation of malicious code gadgets just by swapping around 16-byte blocks. We have shown an instantiation of such a gadget, that maps the \gls{VM}'s stack to a shared page and employs a \gls{ROP} attack to write additional data and execute it as code.

\section{Attack Case Study} \label{sec:attack-scenarios}
With our encryption and code execution gadget from \autoref{sec:oracle}, the attacker gains control over the initially executed code, and is able to insert their own.
In the following, we highlight critical moments in the startup of a \gls{SEV}-secured \gls{VM}, and show how we can use our attack to take it over.

\subsection{Experimental Setup}
Our experiments are performed on a second generation AMD EPYC Processor 7232P.
The firmware of the \gls{SP} is version 24 build 0a~\cite{sp-firmware} (most recent at time of writing). On the \gls{HV} side, we use Linux Kernel 5.6 from the official AMD repository~\cite{amdSEVESKernelRepo} (extended with our attack code) and QEMU~\cite{upstreamQemuRepo} to start the \gls{VM}. QEMU was extended with AMD's \gls{SEV-ES} patches~\cite{amdQemuRepo} and the proposed patches for the secret injection mechanism
from~\cite{jbottomleyencimages} (see \autoref{sec:injecting-secrets}). Inside the \gls{SEV-ES} victim \gls{VM}, we run \gls{OVMF}~\cite{ovmf} and Grub~\cite{gnugrub}. Both were extended with the secret injection mechanism patches from~\cite{jbottomleyencimages}.
We provide our proof-of-concept code alongside with the used software at \url{https://github.com/UzL-ITS/undeserved-trust}.

\subsection{Trust Gap}
For simplicity and scalability, only a small part of the VM's code is attested. In most cases, it suffices to attest a tiny initial code image, which takes owner-supplied secrets to load and decrypt a much larger encrypted disk image, which in turn contains the operating system and the processed data. The operating system can be considered trusted, as the attacker should not be able to modify the encrypted disk image without being in possession of the key.

The primary challenge is bridging the trust gap between the attested initial code image, which is available in plaintext, and the encrypted operating system: The guest owner needs to be able to supply secret information for decrypting the disk image, without an attacker being able to learn those secrets.

If an attacker gains access to a disk encryption key, they also gain unthrottled read/write access to all of the \gls{VM}'s data, even after the \gls{VM} was shut down. They can abuse this access to extract secret data or manipulate the operating system.

\subsection{Securely Injecting Secrets} \label{sec:injecting-secrets}
To address this challenge, SEV offers the \lsecret{} command, which allows the guest owner to inject arbitrary secret values into the VM.

Since there is not yet a standardized toolchain, we focus on the proposed launch flow from~\cite{jbottomleyencimages}, which has some of its patches already merged into the respective upstream repositories. Note that any other possible launch process will also need to bridge the aforementioned trust gap, and will thus be quite similar to the launch flow discussed here.

The proposed launch flow works as follows: The initial attested code image consists of both the \gls{OVMF} UEFI binary and the Grub bootloader, which thus cannot be modified by a malicious HV. The UEFI performs the initial startup, and then transfers control to the bootloader, which in turn unlocks an encrypted disk image and boots the contained Linux kernel.

Both \gls{OVMF} and Grub have been adjusted to respect the secret injection mechanism: At build time, \gls{OVMF} includes a configuration table, which specifies the \gls{GPA} where the \gls{HV} (QEMU) should inject the secret. While preparing the \gls{VM} startup, the \gls{HV} scans the \gls{OVMF} binary, locates the configuration table and subsequently injects the secret at the indicated address. \gls{OVMF} then passes the secret to the Grub bootloader, which uses it to unlock the disk.

\subsection{Leaking the Disk Encryption Key}
Given our attack primitives from \autoref{sec:exploit}, leaking the disk encryption key is quite straightforward. We already know the length of the secret data, since the HV is responsible for receiving the encrypted secret from the guest owner and forwarding it to the SP via \lsecret{}, which expects a public length parameter.
In addition, we know the \gls{GPA} of the secret from \gls{OVMF}'s configuration table. Using our \gls{ROP} chain, we can now inject a small code gadget which loops over the secret data and copies it into shared memory. The code gadget is shown in \autoref{lst:for-loop}.

\begin{figure}[t]
\begin{lstlisting}
5e      pop rsi     ; source address
5f      pop rdi     ; destination address
59      pop rcx     ; n
f3 a4   rep movsb   ; copy n bytes
f4      hlt         ; halt VM
\end{lstlisting}
    \label{lst:for-loop}
    \caption{Code gadget for copying data. The register values are passed via the stack, to minimize the size of our injected code: By using the string copy instruction \texttt{rep movsb}, we can fit the entire gadget into 6 bytes, so we only need a single execution of the \gls{ROP} chain from \autoref{fig:rop-chain}. Since we only leak the secret, we halt the VM after copying the data; however, if we wanted to copy more data, we could replace the \texttt{hlt} instruction by a \texttt{ret} instruction and execute the gadget multiple times with different parameters.}
\end{figure}

We halt the VM after extracting the secret; however, to avoid detection by the guest owner, we could also use our copy gadget to repair the code which we damaged by our block moving approach. Then, potentially after injecting further spy code into the previously encrypted disk image, we resume the boot process.

\section{Mitigations} \label{sec:mitigations}
Our attacks severely undermine the validity of AMD SEV's remote attestation. 
Unfortunately, it is not possible to fully mitigate our attack with the current capabilities
of \gls{SEV-ES}, due to the lack of page remapping protection, which will only be
available with \gls{SEV-SNP} on upcoming 3rd generation EPYC processors. Nonetheless, we discuss changes that could be applied to systems limited to \gls{SEV-ES}, to make exploitation harder.

\subsection{Increasing the Block Size}\label{sec:mitigations:blocksize}
A simple countermeasure, which renders creating an exploit by reordering code blocks much harder, is to increase the minimal size of each measured block. 

Given, e.g., our \gls{OVMF} binary of 3.5 MB, a block size of 16 bytes yields around 230'000 blocks; for a block size of 4 kB, this number shrinks to merely around 900, greatly reducing a malicious HV's ability to find a block ordering that produces meaningful code, although it does not completely mitigate it. Note that \ldata{} already supports large block sizes: To improve performance, the corresponding kernel code tries to process contiguous physical memory blocks, which are chosen as large as possible.

Currently the protocol does not support specifying a certain block size or including it in the launch digest. However, it would be possible to implement a fixed block size without any changes to the protocol, by hardcoding a size of 4 kB (page) or even 2 MB (huge page) directly in the \gls{SP} firmware. This way, the guest owner just needs to require the corresponding firmware version during attestation. Since the version check is already included in the protocol and the \gls{HV} only has to update the \gls{SP} firmware, this countermeasure is rather cheap. Only the \gls{SP} firmware and client applications which verify the launch digest have to be changed.

\subsection{Potential Changes to the Attestation Protocol}
In order to further increase the power of the proposed countermeasure or even fully close the vulnerability, one may also include the physical addresses in the measurement and attestation, or increase the size of the measured blocks and add the block size to the measurement hash. This could be achieved by computing
\begin{equation*}
    h_{i}=\hash(h_{i-1}\concat\mathrm{HPA}_i\concat\mathrm{data}_i)\quad\text{or}
\end{equation*}
\begin{equation*}
    h_{i}=\hash(h_{i-1}\concat\mathrm{block\_size}_i\concat\mathrm{data}_i)
\end{equation*}
on each call $i$ to \ldata{}, for a total of $n$ calls, and submitting the list of addresses or block sizes along the measurement $h_{n}$. Both of these changes require changing the protocol of the remote attestation, since the address list or list of block sizes must be sent along the measurement itself.

However, both approaches are intrinsically limited by the underlying hardware assumptions and address mappings which are in place during virtualization. The HV can still legally reorder physically contiguous 4 kB pages, since it controls the mapping of host physical to guest physical memory addresses through its control over the \gls{NPT}. I.e., the hypervisor is capable of performing page remapping attacks, as already exploited by Morbitzer et al.~\cite{morbitzer2018severed}. Thus, both approaches are limited to assure the correct order within and for the size of one memory page, which can already be achieved with fixing the block size to 4 kB (2 MB) as described in the previous section. The only advantage of including the physical addresses in the measurement is that we can ensure the order inside a  4 kB page, while still allowing 16 byte blocks as the smallest block size.

In conclusion, the attestation process is in need of fixing the loaded binary to addresses within the \emph{guest's} address space: Adding the guest physical address, instead of the host physical address, to the measurement, and assuring that a remapping between guest physical address and host physical address after the initial allocation will be detected by the secure processor, would completely close the vulnerability described in this work. However, SEV-ES does not allow to detect a page remapping and thus only allows for a partial mitigation.

\subsection{Changes in SEV-SNP}
AMD \gls{SEV-SNP}~\cite{sev-snp} is an upcoming extension to \gls{SEV-ES}, which is only supported on the third generation of EPYC processors. As those only become available after the submission deadline, the following is solely based on the documentation~\cite{AMD2020SnPAPI,AMD2020architecture,sev-snp}.

One of the major changes in \gls{SEV-SNP} is the introduction of the \glsfirst{RMP}. The \gls{RMP} is an additional page table, indexed by the \gls{HPA} of a page. It adds several new attributes, mostly for distinguishing \gls{VM} and \gls{HV} pages, such that the HV cannot write to guest pages. In addition, the \gls{RMP} contains the \gls{GPA} of \gls{VM} pages and can be used to ensure a one-to-one mapping between \gls{GPA} and \gls{HPA} to prevent remapping attacks. In contrast to traditional page tables, the \gls{HV} does not have full control over the \gls{RMP}, as it must use hardware- and firmware-mediated ways to access it. 

While the general flow of the launch process does not appear to have changed, there are two essential changes to the \ldata{} command, preventing our attack. The first one implements the idea that we also proposed in \autoref{sec:mitigations:blocksize}: The \gls{HV} must either pass a 4 kB or a 2 MB page to the command (however, 2MB pages are internally treated as multiple 4 kB pages). The second change is the calculation of the launch digest. The hash is now finalized after each call to the launch command and calculated as follows:
\begin{equation*}
     h_{i}=\hash(h_{i-1}\concat\hash(\mathrm{data}_{i})\concat\mathrm{block\_size}_i\concat\dots \concat\mathrm{GPA}_{i} ),
\end{equation*}
where ``$\dots$'' represent additional fields that we omitted for brevity.
Due to hash finalization after each call, the launch digest $h_n$ now reflects the amount of \ldata{} calls used to load the data. In addition, the \texttt{GPA} field
is of special interest, because it includes the memory layout, as it is observed by the guest, in the measurement. According to the documentation, the \gls{GPA} value is computed by the \gls{SP} and also stored in the \gls{RMP}. Furthermore, the page is marked as a guest page. The \gls{GPA} value is enforced, because the hardware page table walker checks that the \gls{GPA} to \gls{HPA} mapping in the \gls{NPT} matches the one in the \gls{RMP}.

\section{Related Work}
Since the initial release of AMD's memory encryption technology, first \gls{SME} and later \gls{SEV}, there has been a wide range of attacks against its security guarantees. Hetzelt et al.~\cite{hetzelt2017security} exploited the  unencrypted register state of the first version of \gls{SEV} to construct simple encryption/decryption oracles. In addition they explored memory replay attacks.

Du et al.~\cite{du2017secure} unveiled the encryption mode, tweak values and the resulting lack of integrity protection of \gls{SME} on Ryzen processors, which is closely related to \gls{SEV} on EPYC processors. They used this knowledge in addition with a network service inside the \gls{VM} to create an encryption oracle on a simulated version of \gls{SEV}, which was not yet available.

Li et al.~\cite{li2019exploiting} used the lack of integrity protection combined with the knowledge of the tweak values to construct encryption as well as decryption oracles. For this, they exploited the fact that \gls{DMA} operations issued by the \gls{VM} are mediated by the \gls{HV} through shared memory pages.

Wilke et al.~\cite{DBLP:conf/sp/WilkeWM020} extended the analysis of the encryption mode and the tweak values to first generation EPYC and EPYC-embedded CPUs, unveiling an updated encryption mode. They showed how to abuse the missing integrity protection combined with knowledge of the tweak values, to bootstrap an encryption oracle from malicious code gadgets, solely by moving around ciphertext blocks in memory. However, the attack does not work on second generation EPYC CPUs, as these feature an enhanced randomization of tweak values. While our attack follows a similar approach of reordering memory blocks, it does not rely on the encryption mode at all, and is therefore also applicable to the currently available second generation EPYC CPUs.

Morbitzer et al.~\cite{morbitzer2018severed} leveraged the \gls{HV}'s control over the \gls{NPT} as well as the page fault side channel to construct a decryption oracle. Similar to Du et al.~\cite{du2017secure}, they require a service running in the \gls{VM}. In their follow-up paper~\cite{DBLP:conf/codaspy/Morbitzer0H19}, they showed how to locate pages containing secrets, like OpenSSH keys, in the \gls{VM}.

Werner et al.~\cite{werner2019severest} showed that it is possible to use the unencrypted register values in the first \gls{SEV} version to reconstruct the code executed in the \gls{VM}. Furthermore, they showed how to use Instruction Based Sampling, a performance counter subsystem, to fingerprint code executed in \gls{SEV-ES} \glspl{VM}.

Radev et al.~\cite{DBLP:journals/corr/abs-2010-07094} described multiple attacks, exploiting insufficient value sanitization at the \gls{HV} to \gls{VM} boundary. For example, they showed how to trick the \gls{VM} into treating arbitrary memory accesses as \gls{MMIO}, as well as into using malicious virtualized cryptographic accelerators provided by the \gls{HV}. In addition, they demonstrated how faking \texttt{cpuid} results can be used to corrupt the \gls{VM} page tables to mark all pages as unencrypted. They then used the unencrypted stack to launch a \gls{ROP} attack, similar to our stage 2 gadget. However, in contrast to our attack, the page table manipulation used by them can be detected by a simple software countermeasure, as described in their paper.

Li et al.~\cite{DBLP:journals/corr/abs-2008-00146} demonstrated that the ``security by crash'' philosophy behind AMD's use of the \gls{ASID} for mapping \glspl{VM} to their memory encryption keys is flawed, as a malicious \gls{HV} can swap the \glspl{ASID} of an attacker \gls{VM} and the victim \gls{VM} to leak limited amounts of data.

Buhren et al.~\cite{buhren2019insecure} explored another attack vector by analyzing the firmware loading mechanism of the \gls{SP}. They discovered a bug allowing them to load customized firmware on the \gls{SP}, breaking the hardware root of trust.

\section{Conclusion}
In this work, we have shown that the current attestation mechanism of SEV has a significant flaw, as it allows the \gls{HV} to reorder blocks of the initially loaded image without influencing the launch measurement, leaving the guest owner unaware of our attack. We have been able to use this vulnerability to redirect execution and inject arbitrary code into the encrypted VM, giving us full control over its execution flow. Moreover, we have shown how this vulnerability in the remote attestation allows us to extract secret data and conduct other attacks, like manipulating the booted operating system.

The attack described in this work undermines the validity of AMD SEV's remote attestation and thus its trustworthiness as a trusted execution environment. Especially, when authenticated and confidential execution in otherwise untrusted environments are required, additional means for verifying the authenticity of the loaded and executed software should be taken into consideration, until the vulnerability is fixed.

We have described possible changes to the firmware of the secure processor, allowing for a simple and reasonably secure mitigation which could be rolled out by means of a firmware update to existing EPYC processors of the first and second generation. However, as argued in \autoref{sec:mitigations}, we do not think that it is possible to completely close this vulnerability with the capabilities of SEV-ES. If the information provided in the SEV-SNP white paper holds, full protection should only become available with the third generation EPYC processors.

\section*{Acknowledgments}
This work was partially supported by the DFG grants 439797619 and 427774779.

\ifAnon
\else

\fi

\ifUsenix
  \bibliographystyle{plain}
  {\footnotesize
  \bibliography{main}
  }
\else
  \bibliographystyle{IEEEtranS}
  {\footnotesize
  \bibliography{IEEEabrv,main}
  }
\fi

\end{document}